# AN OPTIMAL ALGORITHM FOR CONFLICT-FREE COLORING FOR TREE OF RINGS

Einollah Pira

The Business Training Center of Tabriz,Iran
pira_ep2006@yahoo.com

## ABSTRACT

*An optimal algorithm is presented about Conflict-Free Coloring for connected subgraphs of tree of rings. Suppose the number of the rings in the tree is |T| and the maximum length of rings is |R|. A presented algorithm in [1] for a Tree of rings used O(log|T|.log|R|) colors but this algorithm uses O(log|T|+log|R|) colors. The coloring earned by this algorithm has the unique-min property, that is, the unique color is also minimum.*

## 1. INTRODUCTION

A vertex coloring of graph G=(V,E) is an assignment of colors to the vertices such that two adjacent vertices are assigned different colors. A hypergraph H = (V,E) is a generalization of a graph for which hyperedges can be arbitrary-sized non-empty subsets of V. A vertex coloring C of hypergraph H is called conflict-free if in every hyperedge there is a vertex whose color is unique among all other colors in the hyperedge. Suppose the hypergraph H=(V,D) of a graph G=(V,E) be defined as follows: The set of vertices V of H is the same as that of G and the set of hyperedges D consists of all possible subsets of V that induce connected subgraphs of G. Another possible generalization [5] is the following one:

**Definition 1.** A vertex coloring of a hypergraph H=(V,D) is called conflict-free if in every hyperedge e there exists at least one vertex which has a unique color among all other colors used for vertices in that hyperedge.

A vertex coloring of a hypergraph such that the minimum (maximum) color of any vertex of a hyperedge is unique (assigned to only one vertex in this hyperedge) is conflict-free and is called unique-min (resp. unique-max) (confict-free) coloring. The problems of computing a unique-min coloring is equivalent to computing a unique-max coloring since we can replace every color $i$ by $c_{max} - i + 1$, where $c_{max}$ is the maximum color among all vertices [1].

In this paper, first i study unique-min (confict-free) coloring in chain, ring and tree, second, present a new algorithm for a tree of rings.

Conflict-free coloring have various applications. For Example in [2] consider the following scenario: vertices represent base stations of a cellular network interconnected through a backbone. Mobile client connect to the network by radio links and the reception range of each agent is a connected subgraph of the base stations graph. Then it may be desirable that in each agent's range there is a base station transmitting in a unique frequency, in order to avoid interference. The problem of minimizing the number of necessary frequencies is equivalent to Connected Subgraphs Conflict-Free Coloring.

**Related work.** The study of conflict-free coloring was initiated in [2] as a geometric problem with applications to cellular networks. Some of the problems proposed in that paper can be defined as hypergraph conflict-free coloring problems. The algorithm that uses O($\log^2$ n) colors (where n is the number of vertices) is given in [1] about CF-coloring for trees and trees of rings. Some of the problems presented in [2] can be defined as hypergraph conflict-free coloring problems. In [3,4] the conflict-free coloring was studied for grids. In [6] the conflict-free coloring of n points with respect to (closed) disks were studied and were proved a lower bound of $\Omega(\log n)$ colors. In [7] the conflict-free coloring of n points with respect to axis-parallel rectangles were studied. Various other conflict-free coloring problems have been considered in very recent papers [8,12,13,14,15,16,17,18].

The problem becomes more interesting when the vertices are given online by an adversary. For example, at every given time step $i$, a new vertex $v_i$ is given and the algorithm must assign $v_i$ a color such that the coloring is a conflict-free coloring of the hypergraph that is induced by the vertices $V = \{v_1, v_2, ..., v_i\}$. Once $v_i$ is assigned a color, that color cannot be changed in the future. This is an online setting, so the algorithm has no knowledge of how vertices will be given in the future. In [5] there is the online version of conflict-free coloring of a hypergraph. The online version of Connected Subgraphs Conflict-Free Coloring in chains was presented in [8]. Also, in the case of intervals, there are several algorithms [11]. Their randomized algorithm uses $O(\log n \log \log n)$ colors with high probability. Their deterministic algorithm uses $O(\log^2 n)$ colors in the worst case. Recently, randomized algorithms that use $O(\log n)$ colors have been found in [9,10].

## 2. PRELIMINARIES

The topologies i study during this paper are chain, ring, tree and tree of rings. A graph is a ring when all its vertices V are connected in such a way that they form a cycle of length |V|. A tree of rings can be defined recursively in the following manner [18]: it is either a single ring or a ring R attached to a tree of rings T by identifying exactly one vertex of R to one vertex of T. An Example of a tree of rings is displayed in Figure 1.

Algorithm for unique-minimum conflict-free coloring in a chain: in [2] there exists an algorithm that uses $\lfloor \log n \rfloor + 1$ colors for chains. The algorithm for a chain {1,2,…,n} as follows:

step 1: Color vertex $\left\lceil \frac{n}{2^1} \right\rceil$ with color 1

step 2: Color vertices $\left\lceil \frac{n}{2^2} \right\rceil, \left\lceil \frac{n}{2^1} + \frac{n}{2^2} \right\rceil$ with color 2

step 3: Color vertices $\left\lceil \frac{n}{2^3} \right\rceil, \left\lceil \frac{n}{2^2} + \frac{n}{2^3} \right\rceil, \left\lceil \frac{n}{2^1} + \frac{n}{2^3} \right\rceil, \left\lceil \frac{n}{2^1} + \frac{n}{2^2} + \frac{n}{2^3} \right\rceil$ with color 3

…….

step i: Color verices $\left\lceil \frac{n}{2^i} \right\rceil, \ . \ . \ . \ , \left\lceil \frac{n}{2^1} + \frac{n}{2^2} + \frac{n}{2^3} + .... + \frac{n}{2^i} \right\rceil$ with color i

Color i is used only if $\left\lceil \frac{n}{2^i} \right\rceil = 1$, so in fact $\lfloor \log n \rfloor + 1$ colors are used by the algorithm.

For example, if n=8, the coloring is 32313234. It is clearly to see that the coloring is unique-minimum conflict-free coloring.

The above algorithm with a small change can be used to solve the unique-minimum conflict-free coloring in a ring. Pick an arbitrary vertex v and color it with a color 1 (not to be reused anywhere else in the coloring). The remaining vertices form a chain that color with the algorithm described above. This algorithm colors a ring of n vertices with $\lfloor \log(n-1) \rfloor + 2$ colors. For example, if n=8, the coloring is 14342434, where `1' is the first unique color used for v. It is not difficult to see that the coloring is conflict-free: All paths that include v are conflict-free colored, and the remaining graph G–v is a chain of n–1 vertices, so paths of G–v are also conflict-free colored.

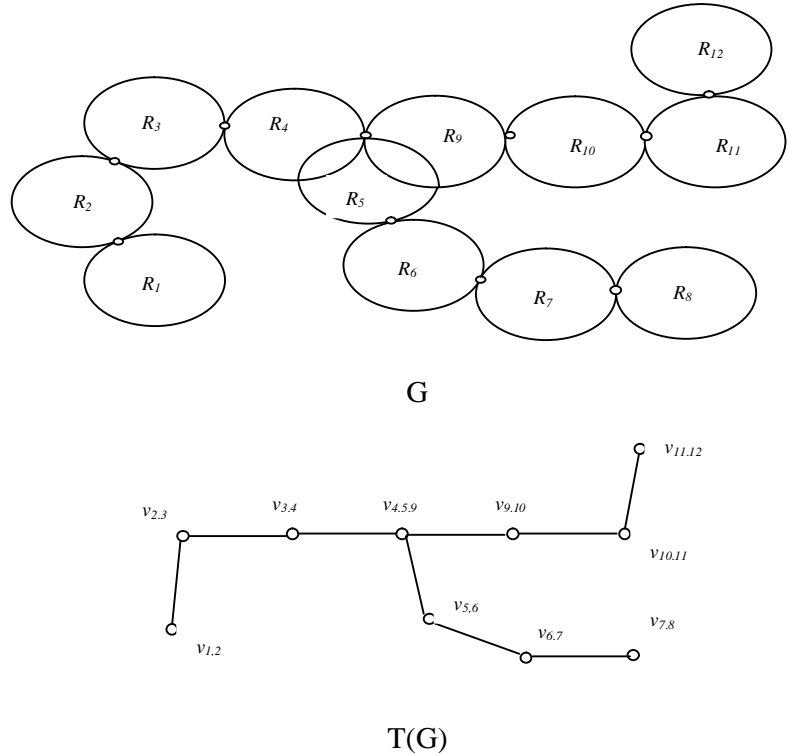

Figure 1. A tree of rings G and the corresponding tree representation T(G).

An important notion for my algorithm is α-separator.

**Definition 2.** An α-separator (α<1) of a graph G=(V,E) is a vertex u the removal of which partitions G to connected components of size at most α|V|.

It is obvious from the above definition that on a general graph an α-separator does not always exist. It is a folklore result that in trees a (1/2)-separator always exists; moreover it can be found in polynomial time [19]. In my algorithm i will often make use of (1/2)-separators.

Algorithm for unique-minimum conflict-free coloring in a tree: in [1] there exists an algorithm that uses $\lfloor \log n \rfloor$ colors for trees. The algorithm for a tree is displayed in Figure 2.

---
**Algorithm 1: Unique-Min Coloring for a Tree**
---
**Input**: a tree T
**Output**: a coloring of vertices of *T*.
**1: Set** $T_1:=T$, $i:=1$.
**2: while** $T_i \neq \phi$ **do**
3:   Find (1/2)-separators on all connected components of forest $T_i$.
4:   Add these separators to set $V_i$.
5:   Color vertices in $V_i$ with color i.
6:   Construct forest $T_{i+1}$ by removing vertices $V_i$ from $T_i$.
7:   Set i:=i+1.
8: **end while**
---

Figure 2.   Algorithm 1

## 3. AN ALGORITHM FOR TREES OF RINGS

In order to present my algorithm for a tree of rings, i will use the notion of tree representation of a tree of rings. Assume a tree of rings G is $R_1, R_2, ..., R_{|T|}$. Let me first describe how to construct such a representation T(G) of a tree of rings G: Connect all vertices together that lied in intersection of rings. An Example of a tree of rings and its tree representation is displayed in Figure 1. The algorithm for a tree of rings is displayed in Figure 3.

---
**Algorithm 2: Unique-Min Coloring for a Tree of Rings**
---
**Input**: a tree of rings G by names $R_1, R_2, ..., R_{|T|}$
**Output**: a coloring of vertices of G
**1:** Construct the tree representation T(G) of the tree of rings G.
**2:** Color the tree T(G) with algorithm 1.
**3:** for i:=1 to |T|-1 do
       Color the vertex in intersection (if exists and before didn't colour) of the rings
       $R_i, R_{i+1}$ by color vertex $v_{i,(i+1)}$ in tree T(G).
   end for
**4:** for i:=1 to |T| do
**5:**      set cm:=a max color of the colored vertices of ring $R_i$.
**6:**      Delete the colored vertices of ring $R_i$ and connect the neighbors of them.
**7:**      Let $R'_i$ denote the resulting cycle.
**8:**      Color cycle $R'_i$ with said algorithm in section 2 by using colors from $\{cm+1,....,cm+\lfloor \log|R'_i| \rfloor +2\}$ .
**9**: end for
---

Figure 3.   Algorithm 2

### 3.1. Analysis of the algorithm

**Lemma 1.** The coloring obtained by Algorithm 2 is a connected-subgraphs unique-min conflict-free coloring.

**Proof**. Assume that C is a path in G. There are two cases for C. **Case 1**: C is part of a ring or a ring itself. if C does not contain the common vertices of the rings, C will be colored in a unique-min way because C colored in line 8 from algorithm 2. if C contains the common vertices of the rings, C will be colored in a unique-min way because the coloring of it start from the max of the colors of the common vertices of the rings (see lines 5,8 from algorithm 2). **Case 2**: C lies on a connected subset of rings, say $R_i,...,R_j$; the corresponding vertices of these rings in T(G), say $v_{i,(i+1)}...v_{(j-1),j}$. Since these vertices of T(G) in line 2 from algorithm 2 are colored in a unique-min way, and each ring $R_k$ in C lies between vertices $v_{(k-1)k}, v_{k,(k+1)}$ that colored in line 8 from algorithm 2, therefore C has been colored in a unique-min way.

**Lemma 2.** The Algorithm 2 uses O(log|T|+log|R|) colors.

**Proof.** The number of colors for coloring T(G) equal log|T|. For coloring the rings, in line 5 from algorithm 2, the maximum of cm's is log|T|, therefore the maximum color is used in line 8 are log|T|+2+log|R|. Thus the Algorithm 2 uses O(log|T|+log|R|) colors.

## 4. CONCLUSIONS

I have presented an optimal algorithm for coloring a tree of rings such that each connected subgraph has a vertex with a unique minimum color. Also i have proved this algorithm uses O(log|T|+log|R|) colors.

## ACKNOWLEDGEMENTS

This research has been supported by the business training center of tabriz,Iran.

## REFERENCES

[1] Georgia, K. & Aris, P. & Katerina, P. (2007) " Conflict-free Coloring for Connected Subgraphs of Trees and Trees of Rings" , In Proc. 11th Panhellenic Conference in Informatics.

[2] Even, G. & Lotker, Z. & Ron, D. & Smorodinsky , S. (2002) "Conflict-free colorings of simple geometric regions with applications to frequency assignment in cellular networks" , SIAM Journal on Computing, 33:94--136. Also in Proceedings of the 43rd Annual IEEE Symposium on Foundations of Compj nu8v uter Science (FOCS).

[3] Bar-Noy, A. & Cheilaris, P. & Lampis, M. & Zachos, S. (2006) "Conflict-free coloring graphs and other related problems", Manuscript.

[4] Cheilaris, P. & Specker, E. & Zachos, S. (2006) Neochromatica. Manuscript.

[5] Bar-Noy, A. & Cheilaris, P. & Smorodinsky, S. (2006) "Conflict-free coloring for intervals: from offline to online", In Proceedings of the 18th Annual ACM Symposium on Parallel Algorithms and Architectures (SPAA 2006), Cambridge, Massachusetts, USA, July 30 – August 2, pages 128 – 137.

[6] Har-Peled, S. & Smorodinsky, S. (2005) "Conflict-free coloring of points and simple regions in the plane", Discrete and Computational Geometry, 34:47—70.

[7] Pach, J. & Toth, G.(2003) "Conflict free colorings. In Discrete and Computational Geometry", The Goodman-Pollack Festschrift, pages 665--671. Springer Verlag.

[8] Alon, N. & Smorodinsky, S. (2006) "Conflict-free colorings of shallow disks" , in Proceedings of the 22nd annual Symposium on Computational Geometry (SCG '06), pages 41-43, New York, NY, USA, ACM Press.

[9] Bar-Noy, A. & Cheilaris, P. & Smorodinsky, S. (2006) "Randomized online conflict-free coloring for hypergraphs", Manuscript.

[10] Chen K. (2006) "How to play a coloring game against a color-blind adversary", In Proceedings of the 22nd Annual ACM Symposium on Computational Geometry (SoCG), pages 44--51.

[11] Fiat A. & Lev M. & Matousek J. & Mossel E. & Pach, J. & Sharir M. & Smorodinsky S. & Wagner U. & Welzl E. (2005) "Online conflict-free coloring for intervals", In Proceedings of the 16th Annual ACM-SIAM Symposium on Discrete Algorithms (SODA), pages 545--554.

[12] Ajwani, D. & Elbassioni, K. & Govindarajan, S. & Ray, S. (2007) "Conflict-free coloring for rectangle ranges using $O(n^{.382+\varepsilon})$ colors", In Proc. 19th ACM Symp. on Parallelism in algorithms and Architectures (SPAA), pages 181-187.

[13] Bar-Noy, A. & Cheilaris, P. & Olonetsky, S. & Smorodinsky, S. (2010) "Online conflict-free coloring for hypergraphs", Combin. Probab. Comput., 19:493-516.

[14] Chen, K. & Kaplan, H. & Sharir, M. (2009) "Online conflict-free coloring for halfplans, congruent disks, and axis-parallel rectangles", ACM Transactions on Algorithms,5(2):16:1-16:24.

[15] Lev-Tov, N. & Peleg, D. (2009) "Conflict-free coloring of unit disks", Discrete Appl. Math., 157(7):1521-1532.

[16] Pach, J. & Tardos, G. (2009) "Conflict-free colorings of graphs and hypergraphs". Combin. Probab. Comput., 18(5):819-834.

[17] Smorodinsky, S. (2009) "Improved conflict-free colorings of shallow discs", manuscript.

[18] Erlebach, T. (2001) "Approximation Algorithms and complexity results for path problems in trees of rings", In Proceedings of the 26th International Symposium on Mathematical Foundations of Computer Science (MFCS 2001), Marianske Lazne, Czech Republic, August 27-31, 2001, pages 351-362.

[19] Erlebach T. & Pagourtzis A. & Potika K. & Stefanakos S, (2003) "Resource allocation problems in multifiber WDM tree networks", In Graph-Theoretic Concepts in Computer Science, 29th International Workshop, WG 2003, Elspeet, The Netherlands, June 19-21, 2003, Revised Papers, LNCS 2880, pages 218-229.

## BIOGRAPHY

- M.S. in Computer Engineering, Sharif University of Technology, Tehran, Iran [2000-2002]
- B.S. in Computer Engineering, Tarbiat Moallam University of Technology, Tehran, Iran [1996-2000]
- Diploma in Math. and Physics, Razi High School, ajabshir, Iran [1992-1996]

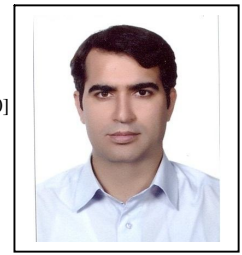